RESEARCH ARTICLE

# Hydrogen Utilization as a Plasma Source for Magnetohydrodynamic Direct Power Extraction (MHD-DPE)

OSAMA A. MARZOUK
College of Engineering, University of Buraimi, Al Buraimi 512, Oman
e-mail: osama.m@uob.edu.om

**ABSTRACT** This study explores the suitability of hydrogen-based plasma in direct power extraction (DPE) as a non-conventional electricity generation method. We apply computational modeling and principles in physics and chemistry to estimate different thermal and electric properties of a water-vapor/nitrogen/cesium-vapor ($H_2O/N_2/Cs$) gas mixture with different levels of cesium (Cs) at a fixed temperature of 2300 K (2026.85 °C). This gas mixture and temperature are selected because they resemble the stoichiometric combustion of hydrogen with air, followed by the addition of the alkali metal element cesium to allow ionization, thus converting the gas mixture into electrically conducting plasma. We vary the cesium mole fraction in the gas mixture by two orders of magnitude, from a minute amount of 0.0625% (1/1600) to a major amount of 16% (0.16). We use these results to further estimate the theoretical upper limit of the electric power output from a unit volume of a high-speed magnetohydrodynamic (MHD) channel, with the plasma accelerated inside it to twice the local speed of sound (Mach number 2) while subject to an applied magnetic field of 5 T (5 teslas). We report that there is an optimum cesium mole fraction of 3%, at which the power output is maximized. Per 1 $m^3$ of plasma volume, the estimated theoretical electric power generation at 1 atm (101.325 kPa) pressure of the hydrogen-combustion mixture is extraordinarily high at 360 $MW/m^3$, and the plasma electric conductivity is 17.5 S/m. This estimated power generation even reaches an impressive level of 1.15 $GW/m^3$ (11500 $MW/m^3$) if the absolute pressure can be decreased to 0.0625 atm (6.333 kPa), at which the electric conductivity exceeds 55 S/m (more than 10 times the electric conductivity of seawater). Our interdisciplinary study combines principles from various fields (gas dynamics, thermodynamics, physics, and chemistry) while analyzing thermochemical and electric properties of weakly-ionized plasma. The study's findings may raise interest in the utilization of hydrogen (particularly green hydrogen) in magnetohydrodynamic direct power extraction (MHD-DPE).

## I. NOMENCLATURE (IN ALPHABETICAL ORDER, GREEK SYMBOLS FIRST), AND [SI UNIT]

$\gamma_{mix}$   Adiabatic index (specific heats ratio, heat capacity ratio, or ratio of the specific heats) of the seeded gas mixture ($H_2O/N_2/Cs$), before ionization [dimensionless].

$\sigma$   Electric conductivity of the plasma [S/m].

$a$   Speed of sound in the plasma, approximated as its value for the seeded gas mixture ($H_2O/N_2/Cs$), before ionization [m/s].

$a_{1,i} - a_{7,i}$   Constant coefficients in the NASA fitting function for computing the normalized (nondimensional) specific heat capacity for a generic gaseous species ($i$), in the high-temperature range (1000-6000 K).

The associate editor coordinating the review of this manuscript and approving it for publication was Tariq Masood.

| | |
|---|---|
| $B$ | Magnetic flux density magnitude of externally applied magnetic field acting on moving plasma within a magnetohydrodynamic channel [T "tesla"]. |
| $C_{p,Cs}$ | Gravimetric (mass-based) specific heat capacity at constant pressure, for cesium vapor [J/kg.K]. |
| $C_{p,H2O}$ | Gravimetric (mass-based) specific heat capacity at constant pressure, for water vapor [J/kg.K]. |
| $C_{p,mix}$ | Gravimetric (mass-based) specific heat capacity at constant pressure, for the seeded gas mixture ($H_2O/N_2/Cs$), before ionization [J/kg.K]. |
| $C_{p,N2}$ | Gravimetric (mass-based) specific heat capacity at constant pressure, for molecular nitrogen [J/kg.K]. |
| $C_{v,mix}$ | Gravimetric (mass-based) specific heat capacity at constant volume, for the seeded gas mixture ($H_2O/N_2/Cs$), before ionization [J/kg.K]. |
| $\mathcal{F}_\sigma$ | Generic function describing the dependence of thermal (equilibrium) alkali-metal-seeded plasma electric conductivity on its temperature, pre-ionization pressure, and pre-ionization chemical composition. |
| M | Mach number [dimensionless]. |
| $M_{Cs}$ | Molecular weight of cesium vapor (Cs) [kg/mol]. |
| $M_{H2O}$ | Molecular weight of water vapor ($H_2O$) [kg/mol]. |
| $M_{mix}$ | Molecular weight of the seeded gas mixture ($H_2O/N_2/Cs$), before ionization [kg/mol]. |
| $M_{N2}$ | Molecular weight of molecular nitrogen ($N_2$) [kg/mol]. |
| $\bar{R}$ | Universal (or molar) gas constant [J/mol.K]. |
| $R_{mix}$ | Specific gas constant of the seeded gas mixture ($H_2O/N_2/Cs$), before ionization [J/kg.K]. |
| $P_V$ | Volumetric power density; electric power output per unit volume of plasma [W/m$^3$]. |
| $p$ | Absolute pressure of the seeded gas mixture ($H_2O/N_2/Cs$), before ionization [Pa]. |
| $T$ | Absolute temperature of the plasma [K]. |
| $u$ | Bulk travel speed of accelerated plasma in a magnetohydrodynamic channel [m/s]. |
| $\vec{X}$ | Vector of the mole fractions of species in a gaseous mixture, which becomes plasma after ionization of some of its alkali metal atoms at a high temperature. |
| $X_{Cs}$ | Mole fraction of cesium vapor, after seeding and before ionization [dimensionless, but expressed here as a percentage, %]. |
| $X_{H2O}$ | Mole fraction of water vapor, after seeding and before ionization [dimensionless, but expressed here as a percentage, %]. |
| $X_{N2}$ | Mole fraction of nitrogen gas, after seeding and before ionization [dimensionless, but expressed here as a percentage, %]. |
| $Y_{Cs}$ | Mass fraction of cesium vapor, after seeding and before ionization [dimensionless]. |
| $Y_{H2O}$ | Mass fraction of water vapor, after seeding and before ionization [dimensionless]. |
| $Y_{N2}$ | Mass fraction of nitrogen gas, after seeding and before ionization [dimensionless]. |

## II. INTRODUCTION
### A. BACKGROUND

The recent success and raised interest in hydrogen production, controlled hydrogen combustion, and hydrogen engines (hydrogen-fueled engines) as zero-emissions alternative energy sources encourage exploring the utilization of hydrogen as a combustion fuel with air, where alkali metal cesium is also added to form an electrically conducting gas (plasma), that can be used as the working fluid in a supersonic linear (straight) magnetohydrodynamic (MHD) channel for direct power extraction (DPE) [1], [2], [3], [4], [5], [6], [7], [8], [9], [10], [11], [12], [13]. The MHD-DPE power generation concept is a non-traditional proven concept (studied and tested, but not commercialized) for extracting direct-current (DC) electric power from high-speed plasma as it passes within an applied magnetic field inside a linear channel, without rotating or moving mechanical elements [14], [15], [16], [17], [18], [19], [20], [21], [22], [23], [24], [25], [26], [27], [28], [29], [30]. The term MHD (magnetohydrodynamics or magneto-hydrodynamics) generally refers to the behavior of electrically conducting continuum fluids (either as gases or as liquids), particularly as they interact with electric or magnetic effects; and it can be encountered and applied under many situations [31], [32], [33], [34], [35], [36], [37], [38], [39], [40], [41], [42].

Although there are plenty of renewable energy sources available for use with mature applicable technologies (such as solar systems and wind systems) to convert renewable energy into clean electricity, green hydrogen (which is hydrogen produced from water electrolysis powered by renewables-based electricity) provides an important feature of being a potential clean energy carrier, rather than an uncontrolled energy flow; and this green hydrogen or any of its derivatives may serve as an alternative fuel replacing fossil fuels in conventional power systems, without the need for complicated carbon capture processes to mitigate the environmental problems involved in fossil fuels [43], [44], [45], [46], [47], [48], [49], [50], [51], [52], [53], [54], [55], [56].

The reaction of hydrogen with oxygen in the air (either through combustion or fuel cells) releases no harmful greenhouse gases (GHGs), particularly carbon dioxide, or air pollutants; but simply releases water vapor. Thus, the expanded utilization of hydrogen as an energy source has significant environmental advantages, through (1) mitigating global warming caused by GHGs, (2) reducing air pollutants emitted during fossil fuel combustion, (3) electrification of transport and urban mobility, (4) improved outdoor air quality in cities, (5) better transition toward green buildings, and (6) facilitating a net-zero carbon economy [57], [58], [59], [60], [61], [62], [63], [64], [65], [66], [67], [68], [69], [70], [71], [72], [73], [74].

Although hydrogen fuel cells, such as the proton exchange membrane (PEM) type, showed successful commercial deployment as standalone electricity generation units or as mobile power units in fuel cell electric vehicles; the current study is concerned with another way of exploiting hydrogen, particularly green hydrogen, in commercial electricity generation through hydrogen-fueled power plants [75], [76], [77], [78]. In this proposed method of stationary large-scale electric power generation, hydrogen is to be combusted with air after a compression stage, leading to a mixture of only water vapor ($H_2O$) and nitrogen ($N_2$) in the ideal case of stoichiometric complete combustion, with no oxygen remaining in the products, and with the temperature being 2300 K (2026.85 °C), that is near the adiabatic flame of hydrogen with air in the absence of special preheating of reactants [79], [80]. Then, a small amount of cesium is added to the combustion products, with the aim of liberating a sufficient number of electrons from some cesium vapor atoms, which makes the resulting gas mixture electrically conducting (thus, making it plasma). Cesium is an alkali metal element, having an atomic number of 55, and is located in period 6 and group 1 (or IA) in the periodic table, below the alkali metal rubidium (Rb) the 5[th] group and above the alkali metal francium (Fr) in the last (7[th]) group [81], [82], [83]. The electron configuration of Cs is $1s^2 2s^2 2p^6 3s^2 3p^6 4s^2 3d^{10} 4p^6 5s^2 4d^{10} 5p^6 6s^1$ or $[Xe]6s^1$. Due to the many electrons shielding the outer valence electron from the nucleus, that outermost electron ($6s^1$) is held very loosely, with weak control by the nucleus over it; so the atom can easily lose that electron [84], [85]. This explains why cesium has the smallest ionization energy among all chemical elements, making it the best seed material to be added for thermal equilibrium ionization (ionization due to high temperature alone, without additional sophisticated techniques) [86]. The ionization energy or ionization potential of cesium refers to the amount of energy required to ionize cesium (to remove the outer electron from a cesium atom), and this is 3.893 eV/atom, or 375.6 kJ/mol [87], [88], [89], [90], [91]. Then, the plasma is accelerated through a convergent-divergent nozzle (a supersonic nozzle) to high speeds to form a supersonic flow within a linear (straight) channel, with the Mach number (ratio of the bulk speed to the local speed of sound) exceeding unity [92], [93], [94], [95], [96], [97]. In that channel, the supersonic plasma travels between two strong external electromagnets, with the magnetic flux density comparable to those used in magnetic resonance imaging (MRI), such that it can reach several teslas (for example, 5 T) [98], [99], [100], [101], [102], [103], [104], [105], [106]. The channel itself has two side electrodes, and due to the interaction of the traveling plasma with the applied magnetic field, an induced direct current (DC) can be collected through the channel electrodes (one positive and one negative) [107], [108]. This type of direct electric power extraction involves no moving mechanical parts, unlike traditional turbogenerators [109], [110], [111], [112], [113], [114]. The DC output electricity can be turned into alternative current (AC) electricity through an inverter, to make it suitable for feeding into an electric grid [115], [116], [117], [118]. A recovery process allows the reuse of the cesium leaving the MHD channel, which is then diverted to the MHD inlet stream. Despite the loss of energy contained in the plasma during the power extraction process and the drop in its temperature at the end of the supersonic MHD channel (this energy is transformed into electricity), its exit temperature and thermal energy are expected to be high enough to allow further extraction of this energy but as heat, rather than as electricity [119], [120], [121]. This takes place by first slowing down the still-supersonic exit plasma to subsonic speeds using a divergent diffuser; and then a heat recovery steam generator (HRSG) is used to transfer heat from the hot exhaust gases to liquid water for boiling and superheating it, to make it superheated steam that is suitable for operating one or more steam turbines, which are connected to conventional mechanical turbogenerators that generate alternating current (AC) electricity, which can be fed into the electric grid as a second stage of a magnetohydrodynamic (MHD) power plant [122], [123], [124], [125], [126]. As in typical steam power cycles, there is a pump stage to pressurize the liquid water before it is boiled, and there is a condenser stage to recover the steam and convert it into liquid water again for reuse.

Fig. 1 lists the 11 main components of a hypothetical (proposed) hydrogen-fueled magnetohydrodynamic (MHD) plasma power plant, showing the input chemical materials and the output electric power at the two generation stages.

### B. OBJECTIVES AND MODELLING APPROACH

The design of a hydrogen magnetohydrodynamic (MHD) power plant is complicated, with coupling among multiple components in different fields of engineering, science, and technology; such as fluid mechanics, gas dynamics, thermodynamics, heat transfer, plasma physics, electromagnetism, turbomachinery, and combustion [127], [128], [129], [130], [131], [132], [133], [134], [135], [136], [137], [138], [139], [140], [141], [142], [143], [144], [145], [146]. The use of computational procedures aided with analytical modeling for reduced idealized configurations is important in studying such a system, as a preliminary phase to obtain rough estimates.

In the current study, we aim at estimating approximate upper limits for the electric power output from a supersonic hydrogen-based plasma. To this end, we combine three submodels. The first submodel is an analytical single-step expression for a simplified one-dimensional channel problem, that relates the maximum volumetric electric power output to the plasma's electric conductivity, the plasma's bulk

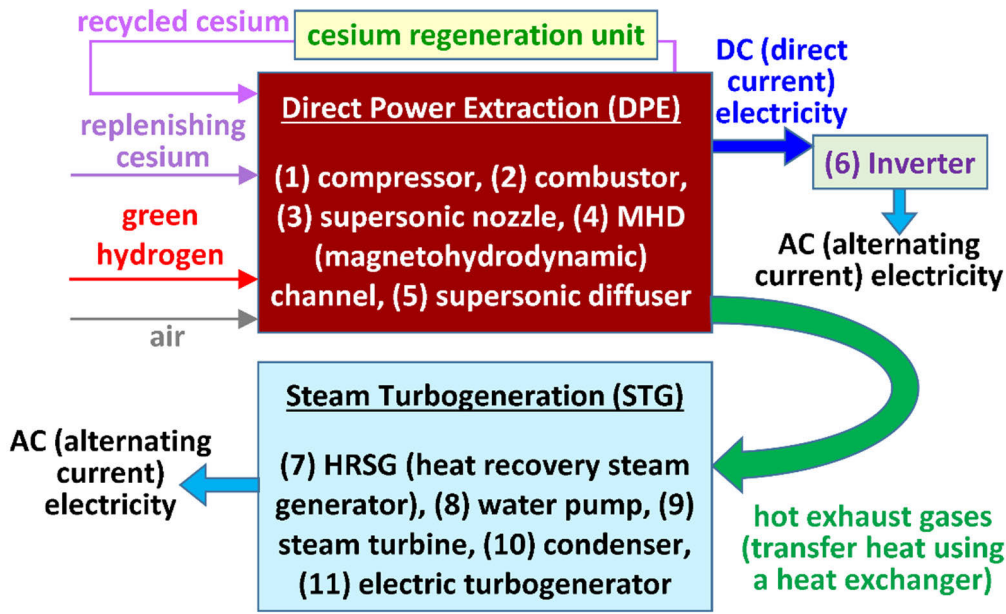

**FIGURE 1.** Illustration of the 11 main elements and the energy transfers in a proposed hydrogen-fueled magnetohydrodynamic power plant.

travel speed, and the applied external magnetic field flux density. The second submodel is a detailed procedure to compute the electric conductivity of a local zone of plasma given its temperature, and the pressure and chemical composition (including the seed alkali metal) of its pre-ionization gas mixtures. The electric conductivity of the plasma directly influences its ability to conduct electric current internally between the channel electrodes, and impacts the effectiveness of electric power extraction and electric current flow to a load connected externally to these electrodes by a solid conductor. The third submodel implements laws of gaseous mixtures to find important thermodynamic properties of the $H_2O/N_2/Cs$ gas mixture that results from adding the seed vapor cesium (Cs) to the hydrogen-air stoichiometric combustion products ($H_2O$ and $N_2$) which turn into plasma after partial ionization of the cesium vapor atoms. In this submodel, we find the specific gas constant of this mixture and its adiabatic index. These two thermodynamic properties allow computing the speed of sound in this gas mixture (for a given temperature). With the Mach number fixed at a mild supersonic value of two, and the external magnetic flux density also fixed at a reasonable selected value of 5 T, then the theoretical electric power output per unit volume of the MHD plasma channel can be estimated. The temperature is fixed at 2300 K (near the maximum "adiabatic" value for air-hydrogen combustion). The main independent variable in the current study is the mole fraction of the seeded cesium. It is varied over a wide range and its effect on the electrothermal properties of the plasma is explored. The ambient absolute pressure of the seeded hydrogen-combustion gas mixture ($H_2O/N_2/Cs$) is considered at three values, namely a moderate pressure of 1 atm (101.325 kPa), a higher pressure of 16 atm, and a lower pressure (partial vacuum) of 1/16 atm (0.0625 atm or

6.3328125 kPa). Therefore, a total of 39 data points (39 operational possibilities) are selected in the current study, and an optimum cesium level is sought, as well as a recommendation for the pressure. The numerical results are presented graphically after proper data processing, and the variation of various MHD plasma performance quantities with the cesium mole fraction or the pressure is examined in view of optimizing the electric power extraction [147], [148], [149], [150], [151], [152], [153], [154], [155].

## III. RESEARCH METHOD

This section is dedicated to explaining the three submodels used in the current computational study, and how we obtained the reported numerical results.

### A. SUBMODEL 1: VOLUMETRIC POWER OUTPUT

For an electric conductivity ($\sigma$) of unfirm one-dimensional plasma within a magnetohydrodynamic (MHD) channel with a rectangular cross-section, an applied magnetic flux density magnitude ($B$), and a bulk speed of the plasma as a fluid ($u$); the optimum electric power output to an external matched load (a load whose electric resistance maximizes its power consumption) per unit volume of the plasma ($P_V$) is [156], [157], [158], [159]

$$P_V = 0.25 \sigma u^2 B^2 \qquad (1)$$

We recently provided a detailed proof of this analytical expression in another study [160]. These details are not repeated here; only the final expression is listed. Further supportive references include [161], [162], and [163].

It should be noted that the traveling speed ($u$) of the plasma through the MHD channel can be expressed as the product of the Mach number (M) and the local speed of sound ($a$) in the

plasma, thus

$$u = Ma \quad (2)$$

Therefore, (1) can be rewritten in an alternative form as

$$P_V = 0.25 \sigma M^2 a^2 B^2 \quad (3)$$

When the unit of the plasma electric conductivity is S/m, the unit of the speed of sound is m/s, and the unit of the magnetic flux density is T (tesla), the volumetric electric power output becomes in W/m$^3$. An additional division by $10^6$ is needed to convert this to the more convenient unit of MW/m$^3$, which is equivalent to W/cm$^3$.

Equation (3) suggests linear correlation between the electric power output and the electric conductivity, while suggesting stronger quadratic correlation between the electric power output and either the speed of sound or the magnetic flux density. In the current study, the Mach number is fixed at a selected value of 2, while the magnetic flux density is fixed at a selected value of 5 T. Thus, (3) can be customized to our study as

$$P_V \left[\frac{MW}{m^3}\right] = 2.5 \times 10^{-5} \sigma \left[\frac{S}{m}\right] a \left[\frac{m}{s}\right]^2 \quad (4)$$

The use of (4) requires the ability to compute the plasma electric conductivity ($\sigma$) and the speed of sound in plasma ($a$). The submodels for these two plasma properties are described in the next two subsections, respectively.

### B. SUBMODEL 2: ELECTRIC CONDUCTIVITY OF PLASMA

The electric conductivity ($\sigma$) of the plasma that is formed as a result of equilibrium (thermal) ionization of a mixture of gaseous species including a seeded amount of alkali metal, at a given temperature ($T$) and at a given pressure ($p$) can be described as the following generic expression:

$$\sigma = \mathcal{F}_\sigma \left(\vec{X}, T, p\right) \quad (5)$$

where the symbol ($\vec{X}$) refers to the list (or vector) of mole fractions of the gaseous species contained in the mixture. The pressure here is expressed as an absolute value (measured from the absolute zero point, not from the ambient atmospheric point), and the temperature here is also expressed as an absolute value (measured from the absolute zero point, in kelvins) [164], [165], [166], [167], [168], [169].

In the current study, the gas mixture that becomes plasma after ionization has three chemical species, which are: water vapor (H$_2$O) from burning the hydrogen, nitrogen (N$_2$) from the oxidizer air, and cesium vapor (Cs) as seeded easily-ionizable material. Thus, a customized form of (5) for the current study is

$$\sigma = \mathcal{F}_\sigma \left(X_{H2O}, X_{N2}, X_{Cs}, T, p\right) \quad (6)$$

where ($X_{H2O}$) is the mole fraction of water vapor, ($X_{N2}$) is the mole fraction of molecular nitrogen, and ($X_{Cs}$) is the mole fraction of vaporized cesium.

Because the sum of the three mole fractions must be exactly one, one of the three mole fractions is dependent on the two other mole fractions and thus does not need to be specified. Therefore, the mole fraction of water vapor can be eliminated as

$$X_{H2O} = 1 - X_{N2} - X_{Cs} \quad (7)$$

In addition, because the hydrogen-air combustion is assumed to be stoichiometric (no excess air, no excess hydrogen), another relation exists that puts a constraint on how the mole fractions can vary. In our study, air is treated as a mixture of molecular oxygen (O$_2$) and molecular nitrogen (N$_2$) with the mole fractions of 21% (0.21) for oxygen and 79% (0.79) for nitrogen; thus for 1 mole of oxygen, air is treated as having 3.762 moles of nitrogen ( 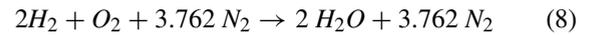[170], [171], [172], [173], [174], [175]. The idealized complete combustion of hydrogen in air becomes

$$2H_2 + O_2 + 3.762\, N_2 \rightarrow 2\, H_2O + 3.762\, N_2 \quad (8)$$

Thus, the molar ratio of nitrogen to water vapor in the pre-ionization gas mixture is (3.762/2) or 1.881. Therefore, the pre-ionization seeded gas mixture has the following condition for its chemical composition

$$X_{N2} = 1.881\, X_{H2O} \quad (9)$$

Using (9) in (7) gives

$$X_{H2O} = 1 - 1.881\, X_{H2O} - X_{Cs} \quad (10)$$

Separating terms with different mole fractions gives

$$2.881\, X_{H2O} = 1 - X_{Cs} \quad (11)$$

Thus, the mole fraction of hydrogen can be eliminated (can be deduced from the mole fraction of cesium) as

$$X_{H2O} = \frac{1 - X_{Cs}}{2.881} \quad (12)$$

Using the above (12) in (9) shows that also the mole fraction of nitrogen does not need to be specified, but can be inferred directly from the mole fraction of cesium as

$$X_{N2} = 1.881 \frac{1 - X_{Cs}}{2.881} = 0.6528983\,(1 - X_{Cs}) \quad (13)$$

From (12, 13), (6) for computing the plasma electric conductivity can be further customized to our study by showing that the electric conductivity requires specifying only the temperature (which is fixed as 2300 K in our study), the pressure of seed gas mixture (H$_2$O/N$_2$/Cs), and the mole fraction of cesium in this pre-ionization gaseous mixture. Thus, (6) is reduced to

$$\sigma = \mathcal{F}_\sigma\, (X_{CS}, T, p) \quad (14)$$

It should be noted that the symbol ($\mathcal{F}_\sigma$) does not refer to a single-step function, but rather refers to a procedure of multiple steps with several physical constants and intermediate quantities, which ultimately leads to the numerical estimation of the electric conductivity as a combined result of two effects: (1) the scattering of electrons due to collisions with neutral atoms, (2) the scattering of electrons due to the presence of positively-charged cesium ions and other negatively-charged electrons. We recently provided a detailed mathematical description and validation of this procedure in another study [176]. These details are not repeated here. Further supportive references include [177], [178], [179], [180], and [181].

For solid conductors, which have well-defined geometric dimensions (length and cross-section area), the ability to conduct electric current is expressed as a resistance value (in ohms, Ω). However, for fluid conductors (liquids or gases), such geometric constraints are not always applicable; and the electric conductivity (rather than resistance) serves as a more proper alternative to quantify their ability to conduct electric current densities (electric current per unit area) under a given electric field, with the unit being S/m or 1/(Ω.m). The procedure implied by (14) provides the electric conductivity of plasma in the SI (système international, international system) unit of (S/m) such that it is ready for incorporation in (4) for computing a theoretical volumetric power density of the MHD channel. It may be useful to compare this computed plasma electric conductivity to a reference point that is easily understood; just to have a better feeling of the strength of such electric conductivity. For this, we selected the value of 5 S/m as an estimate for the electric conductivity (or EC) for seawater in oceans, which approximately corresponds to a salinity concentration of 35000 ppm (parts per million, by mass) or 3.5% (by weight); and these are roughly 100 times more than the electric conductivity and salinity level of drinkable water [182], [183], [184], [185], [186], [187], [188], [189], [190].

### C. SUBMODEL 3: SPEED OF SOUND

The remaining quantity that is needed for estimating the theoretical maximized electric power output from the hydrogen-fueled MHD is the speed of sound in the plasma. Due to the weak ionization level (only a small part of cesium atoms becomes cesium ions, and cesium itself is generally a small component of the entire pre-ionization mixture), this speed of sound is computed for the pre-ionization mixture of ($H_2O$/$N_2$/Cs). Among all the 39 cases examined here, the highest fraction of either the cesium ions ($Cs^+$) or the free electrons ($e^-$) in the plasma (after ionization) is only 0.0792% (0.000792), with neutral molecules ($H_2O$/$N_2$/Cs) accounting for the overall majority of 99.8416% (0.998416). The arithmetic mean (simple average) mole fraction of post-ionization electrons or cesium ions over all the 39 cases is much less as 0.0147% or 0.000147, which is very close to zero (with 99.9706% or 0.999706 of the plasma being neutral species, very close to one); and this justifies ignoring the role of cesium ions in the speed of sound calculations within the plasma.

As for ideal gases, the speed of sound ($a$) in the plasma is approximated through the following relation with the absolute temperature ($T$), the specific gas constant of the $H_2O$/$N_2$/Cs mixture ($R_{mix}$), and the adiabatic index of that mixture ($\gamma_{mix}$) as [191], [192]

$$a \cong \sqrt{\gamma_{mix} R_{mix} T} \quad (15)$$

The adiabatic index of the pre-ionization gas mixture is defined as the ratio of the specific heat capacity at constant pressure for the mixture ($C_{p,mix}$) to the specific heat capacity at constant volume for the mixture ($C_{v,mix}$); which is mathematically expressed as

$$\gamma_{mix} = \frac{C_{p,mix}}{C_{v,mix}} \quad (16)$$

The specific heat capacity at constant pressure is the sum of the specific heat capacity at constant volume and the specific gas constant

$$C_{v,mix} = C_{p,mix} - R_{mix} \quad (17)$$

Therefore, the specific heat at constant volume for the mixture ($C_{v,mix}$) can be eliminated from (16), such that knowledge of the specific heat capacity at constant pressure for the mixture ($C_{p,mix}$) along with the specific gas constant for the mixture ($R_{mix}$) becomes sufficient to obtain the adiabatic index for the mixture ($\gamma_{mix}$), as

$$\gamma_{mix} = \frac{C_{p,mix}}{C_{p,mix} - R_{mix}} \quad (18)$$

Following principles of chemical mixtures, the specific gas constant of the $H_2O$/$N_2$/Cs mixture ($R_{mix}$) can be computed from the universal (or molar) gas constant ($\bar{R}$) and the molecular weight of the mixture ($M_{mix}$) as [193], [194]

$$R_{mix} = \frac{\bar{R}}{M_{mix}} \quad (19)$$

The universal gas constant is 8.314462618 J/mol.K [195].

The molecular weight of the mixture ($M_{mix}$) is computed as a weighted average of the molecular weights of the species in the mixture (weighted by the mole fraction), thus

$$M_{mix} = X_{H2O} M_{H2O} + X_{N2} M_{N2} + X_{Cs} M_{Cs} \quad (20)$$

The molecular weights of water vapor, molecular nitrogen, and cesium vapor were taken from the NSIT Chemistry WebBook database (by the United States National Institute of Standards and Technology), as shown in Table 1.

**TABLE 1.** Molecular weights of the individual components in the pre-ionization GAS mixture.

| Species ↓ | Molecular Weight [kg/mol] | Reference |
|---|---|---|
| Water ($H_2O$) | 0.018015.3 | [196] |
| Nitrogen ($N_2$) | 0.0280134 | [197] |
| Cesium (Cs) | 0.1329054519 | [198] |

The specific heat capacity at constant pressure for the $H_2O$/$N_2$/Cs mixture ($C_{p,mix}$) is computed as a weighted average (weighted by the mass fractions) of the specific

**TABLE 2.** Specific GAS constants and coefficients of the fitting functions to compute the specific heat capacity at constant pressure for the individual components in the pre-ionization GAS mixture.

| Species → | Water Vapor ($H_2O$) | Molecular Nitrogen ($N_2$) | Cesium Vapor (Cs) |
|---|---|---|---|
| $R_i$ | 461.52230 | 296.80305 | 62.55923 |
| $a_{1,i}$ | 1.034972096e06 | 5.877124060e05 | 6.166040900e06 |
| $a_{2,i}$ | −2.412698562e03 | −2.239249073e03 | −1.896175522e04 |
| $a_{3,i}$ | 4.646110780e00 | 6.066949220e00 | 2.483229903e01 |
| $a_{4,i}$ | 2.291998307e−03 | −6.139685500e−04 | −1.251977234e−02 |
| $a_{5,i}$ | −6.836830480e−07 | 1.491806679e−07 | 3.309017390e−06 |
| $a_{6,i}$ | 9.426468930e−11 | −1.923105485e−11 | −3.354012020e−10 |
| $a_{7,i}$ | −4.822380530e−15 | 1.061954386e−15 | 9.626500908e−15 |

heat capacities at constant pressure for the three individual gaseous species, or

$$C_{p,mix} = Y_{H2O}C_{p,H2O} + Y_{N2}C_{p,N2} + Y_{Cs}C_{p,Cs} \quad (21)$$

The mass fractions ($Y_{H2O}$, $Y_{N2}$, $Y_{Cs}$) of the three species ($H_2O$, $N_2$, Cs), respectively, are computed as

$$Y_{H2O} = \frac{X_{H2O}M_{H2O}}{M_{mix}}, Y_{N2} = \frac{X_{N2}M_{N2}}{M_{mix}}, Y_{Cs} = \frac{X_{Cs}M_{Cs}}{M_{mix}} \quad (22)$$

The specific heat capacity at constant pressure for each of the three species ($H_2O$, $N_2$, Cs) is modeled here as a temperature-dependent function according to the NASA 9-coefficient formulation for thermodynamic properties, keeping only the high-temperature range (from 1000 K to 6000 K) that is of interest in our problem of magnetohydrodynamic direct power extraction [199], [200], [201], [202], [203], [204]. It should be noted that elemental cesium normally vaporizes at a boiling point near 940 K and fuses at a low boiling point of approximately 302 K, which is near ordinary room temperatures [205], [206], [207], [208], [209], [210]. Therefore, although cesium is a solid metal at a room temperature of 25 °C (but close to melting), it is treated here as a vaporous species. For a general gaseous species ($i$), the nondimensional ratio of the specific heat capacity at constant pressure to the specific gas constant ($C_{p,i}/R_i$) can be expressed as a seven-term nonlinear pseudo-polynomial function in the absolute temperature, with power terms up to quartic, as

$$C_{p,i}/R_i = a_{1,i}T^{-2} + a_{2,i}T^{-1} \\ + a_{3,i} + a_{4,i}T + a_{5,i}T^2 + a_{6,i}T^3 + a_{7,i}T^4 \quad (23)$$

The specific gas constant ($R_i$) for such a generic species is the result of dividing the universal gas constant ($\bar{R}$) by the molecular weight ($M_i$), or

$$R_i = \frac{\bar{R}}{M_i} \quad (24)$$

Table 2 lists the specific gas constants and the coefficients for the normalized (nondimensional) specific heat capacity for the three species of the seeded gas mixture for a hydrogen-fueled magnetohydrodynamic (MHD) channel.

### D. INPUT CONDITIONS

The main independent variable in the current study is the mole fraction of cesium (Cs) in the pre-ionization seeded gaseous mixture, which is denoted by the symbol ($X_{Cs}$). The value of this design variable is changed by two orders of magnitude; and the resulting changes in various dependent variables related to the formation of the magnetohydrodynamic (MHD) plasma are computed in response to the change of the cesium mole fraction. Due to the wide variation in the cesium mole fraction, a logarithmic spacing (log scale) is more favored over an ordinary equal spacing (linear scale), to avoid smearing the lower end of the examined range (at smaller cesium mole fractions). Specifically, the selected values of the cesium mole fraction start from a lower bound of 0.0625% (0.000625), which is doubled successively to form additional values, until an upper bound value of 16% (0.16) is reached. This range is so wide that its extreme values might not be practically very feasible, since a fraction of 16% cesium is too large when it is considered as a minor seed additive, and a fraction of 0.0625% is so small for the cesium ionization to have appreciable impact. Despite this, this wide range of cesium mole fraction is retained as it helps explore the possible behavior of the MHD-DPE (magnetohydrodynamic direct power extraction) when green hydrogen is used as a clean combustion fuel. The unequal spacing algorithm leads to nine data points, with nine cesium mole fractions; namely 0.0625%, 0.125%, 0.25%, 0.5%, 1%, 2%, 4%, 8%, and 16%. Four additional intermediate points were added at the cesium mole fractions of 3%, 5%, 6%, and 7% because large changes were observed over this region in either the plasma electric conductivity or the volumetric power density; where a peak point occurs. Thus, a total of 13 values of the cesium mole fraction are explored in our study. Furthermore, three values for the pre-ionization pressure are also covered; namely a moderate value of 1 atm (101.325 kPa), a lower value of 0.0625 atm (6.3328125 kPa), and a high value of 16 atm (1621.2 kPa or 1.6212 MPa). These selected values of pressures make the pressure variable explored over two orders of magnitude similar to the cesium mole fraction, with the ratio of the maximum-to-minimum ratio being 256 ($2^8$) for either variable. Thus, a total of 39 data points (39 possible operational conditions defined by the pair of cesium mole fraction and pre-ionization pressure). The temperature is fixed at 2300 K in all these cases. It is useful to mention here that although the pressure slightly increases during ionization (due to the liberated electrons from some cesium atoms, which appear as additional particles with their additional partial pressure), the weak degree of cesium ionization makes this increase so small that the pressure may be approximated as constant during the ionization process. Thus, the pressure of plasma (composed of $H_2O/N_2/Cs/Cs^+/e^-$) is nearly the same as the pressure of the pre-ionization non-electrically-conducting seeded gas mixture (composed of $H_2O/N_2/Cs$).

**TABLE 3.** Mole fractions corresponding to the 13 considered levels of cesium seeding.

| Cesium Vapor Mole Fraction ($X_{Cs}$) | Water Vapor Mole Fraction ($X_{H2O}$) | Molecular Nitrogen Mole Fraction ($X_{N2}$) |
|---|---|---|
| 0.062500% (0.000625, 1/1600) | 34.688476% | 65.249024% |
| 0.125000% (0.00125, 1/800) | 34.666782% | 65.208218% |
| 0.250000% (0.0025, 1/400) | 34.623395% | 65.126605% |
| 0.500000% (0.005, 1/200) | 34.536619% | 64.963381% |
| 1.000000% (0.01, 1/100) | 34.363068% | 64.636932% |
| 2.000000% (0.02, 1/50) | 34.015967% | 63.984033% |
| 3.000000% (0.03, 1/33.333333) | 33.668865% | 63.331135% |
| 4.000000% (0.04, 1/25) | 33.321763% | 62.678237% |
| 5.000000% (0.05, 1/20) | 32.974662% | 62.025338% |
| 6.000000% (0.06, 1/16.666667) | 32.627560% | 61.372440% |
| 7.000000% (0.07, 1/14.285714) | 32.280458% | 60.719542% |
| 8.000000% (0.08, 1/12.5) | 31.933356% | 60.066644% |
| 16.000000% (0.16, 1/6.25) | 29.156543% | 54.843457% |

Table 3 lists the mole fractions of the pre-ionization gas mixture $H_2O/N_2/Cs$ for the 13 chemical compositions considered in the current study. These compositions remain the same for the three values of the pressure considered.

## IV. RESULTS

The results are divided into four subsections. In each subsection, we present the influence of the two controllable input variables (the cesium level and, if applicable, the pressure) in the study on one or more performance variables (dependent output) in relation to the hydrogen-based plasma formation and the prospective electric power generation in a magnetohydrodynamic linear channel (MHD generator).

The output (dependent) variables to be demonstrated are the specific gas constant ($R_{mix}$), the adiabatic index ($\gamma_{mix}$), and the specific heat capacity at constant pressure ($C_{p,mix}$) of the $H_2O/N_2/Cs$ gas mixture that becomes plasma after partial ionization of cesium; the approximated speed of sound in the plasma ($a$); and electric conductivity of the formed plasma ($\sigma$); and the estimated theoretical electric power output per unit volume of plasma ($P_V$).

The results are presented in the form of response curves, with the numerical values of the output variables superimposed over each curve for high reporting precision.

### A. GAS CONSTANT AND ADIABATIC INDEX

The specific gas constant ($R_{mix}$) and the adiabatic index ($\gamma_{mix}$) of the cesium-seeded combustion gas products of hydrogen-air combustion are shown in Fig. 2.

Because a cesium molecule (which is also a cesium atom) is heavier than the molecule of water vapor (ratio of their masses 7.377) and heavier than the molecule of molecular nitrogen (ratio of their masses 4.744); as the level of cesium increases in the gas mixture, the overall molecular weight increases, and thus the mixture's specific gas constant declines. In the limiting case of vanishing cesium, the gas mixture resulting from the complete simple combustion of one mole of hydrogen in air (represented as a mixture of 21% oxygen and 79% nitrogen, by mole fraction) becomes only water vapor and molecular nitrogen with the composition of (2 $H_2O$, 3.762 $N_2$). The mole fractions of this mixture are 2/5.762 or 21% for $H_2O$, and 3.762/5.762 or 79% for $N_2$, and the molecular weight is 0.02454304 kg/mol, leading to a specific gas constant of 338.7707 J/kg.K. With a cesium mole fraction of 2% (0.02), the decrease in the specific gas constant relative to the cesium-free mixture is 27.4876 J/kg.K, or 8.114% of the cesium-free value. This relative change is approximately halved and becomes 4.229% for a cesium mole fraction of 1% (0.01).

The reciprocal of (18) for the adiabatic index ($\gamma_{mix}$) of the $H_2O/N_2/Cs$ gas mixture is

$$\frac{1}{\gamma_{mix}} = \frac{C_{p,mix} - R_{mix}}{C_{p,mix}} = 1 - \frac{R_{mix}}{C_{p,mix}} \quad (25)$$

The reciprocal of the above (25) gives an alternate but equivalent form for ($\gamma_{mix}$) as

$$\gamma_{mix} = \frac{1}{1 - R_{mix}/C_{p,mix}} \quad (26)$$

From the above (26), it can be shown that when the specific gas constant of the $H_2O/N_2/Cs$ gas mixture ($R_{mix}$) drops at the same rate as the specific heat capacity at constant pressure of that mixture ($C_{p,mix}$), then the mixture's adiabatic index remains unchanged. However, if ($R_{mix}$) drops slower than ($C_{p,mix}$), the ratio ($R_{mix}/C_{p,mix}$) grows, and thus the denominator in (26) for ($\gamma_{mix}$) shrinks, ultimately causing ($\gamma_{mix}$) to increase. This is what we found to happen in our study, where the nondimensional fractional ratio ($R_{mix}/C_{p,mix}$) increases from 0.196136 (computed as 337.8384/1722.472) at $X_{CS}$ = 0.0625 to 0.212392 (computed as 198.5258/934.712) at $X_{CS}$ = 16. Thus, the increase in ($R_{mix}/C_{p,mix}$) over the entire range of ($X_{CS}$) is 8.288% (relative to the starting value of 0.196136 at $X_{CS}$ = 0.0625). The decline in ($C_{p,mix}$) as the cesium mole fraction increases is shown in Fig. 3.

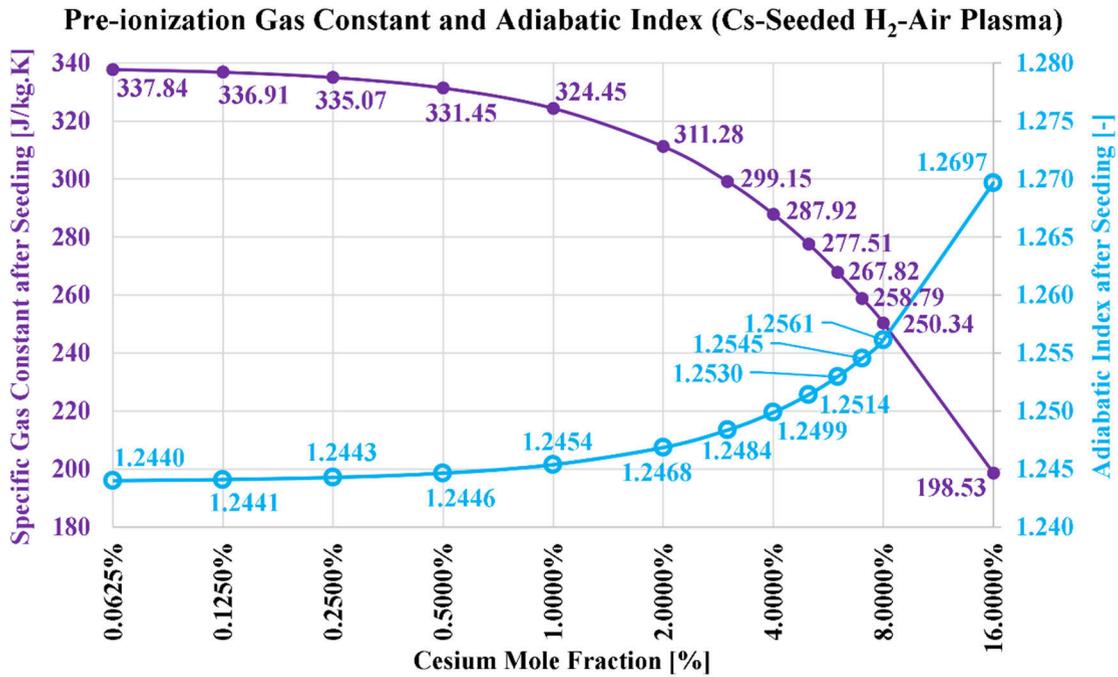

**FIGURE 2.** Response curves of the specific gas constant ($R_{mix}$) and the adiabatic index ($\gamma_{mix}$) of the $H_2O/N_2/Cs$ gas mixture.

It is useful to add here that despite the increase in the ratio ($R_{mix}/C_{p,mix}$) that caused ($\gamma_{mix}$) to increase; such an increase in ($R_{mix}/C_{p,mix}$) is not large, and consequently the increase in ($\gamma_{mix}$) is also small, varying slowing by only 2.064% from its starting value of 1.243991 at $X_{CS}$ = 0.0625% to 1.269668 at $X_{CS}$ 16; thus varying by a difference of only 0.025677. Therefore, the adiabatic index ($\gamma_{mix}$) of the $H_2O/N_2/Cs$ gas mixture is not very sensitive to the level of cesium seed, and ($\gamma_{mix}$) remains within a narrow range near 1.25. This behavior of weak change in ($\gamma_{mix}$) indicates that both types of the specific heat capacity ($C_{p,mix}$, $C_{v,mix}$) decline as the cesium concentration increases, but their rates of decline are similar, causing them to maintain a ratio near 1.25.

We clarify that the specific heat capacity at constant pressure, the specific gas constant, and the adiabatic index are independent of the pressures. Thus, there is a single response curve for each of these three properties as a function of the cesium mole fraction (rather than one curve for each pressure value).

### B. SPEED OF SOUND

As described earlier in (15), the approximated speed of sound in the plasma formed by the partial ionization of cesium in the hydrogen-air combustion plasma depends on the adiabatic index, the specific gas constant, and the temperature. Because the temperature is fixed at 2300 K, and because the previous subsection showed that the adiabatic index varies slightly around a value of 1.25; it is the specific gas constant that primarily dictates how the speed of sound is affected by the level of cesium.

Fig. 4 confirms this remark, by showing that the response curve of the speed of sound resembles qualitatively that of the specific gas constant, with a monotonic decline by 22.556% from 983.166 m/s (3539.4 km/h) at $X_{CS}$ = 0.0625 761.408 m/s (2741.1 km/h) at $X_{CS}$ = 16. In the limiting case of unseeded $H_2O/N_2$ mixture (with mole fractions 21%/79%, respectively), we found that the speed of sound is 984.486 m/s (3,544.1 km/h). With a cesium mole fraction of 2% (0.02), the decrease in the speed of sound relative to the cesium-free mixture is 39.667 m/s, or 4.029% of the cesium-free value. Therefore, the effect of cesium on this property remains relatively small up to a level of 2% mole fraction. This relative change is even smaller as only 2.079% for a cesium mole fraction of 1% (0.01). As a reference benchmarking value, the speed of sound in air at normal conditions (15 °C and 1 atm) is approximately 340 m/s (1224 km/h) [211], [212], [213], [214], [215].

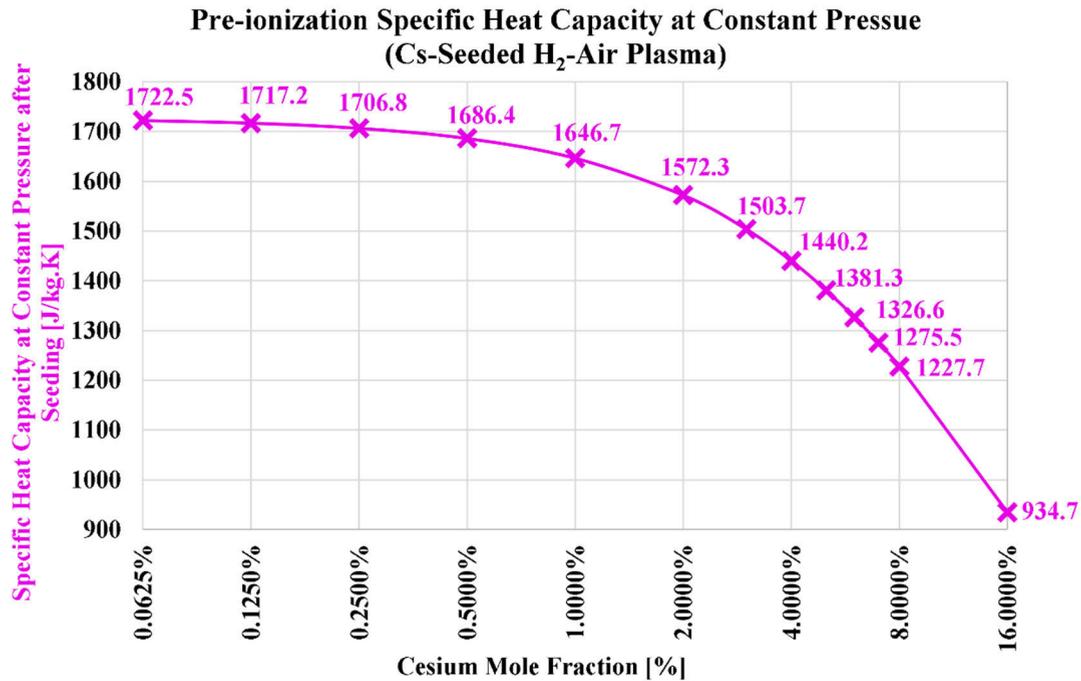

**FIGURE 3.** Response curve of the specific heat capacity at constant pressure ($C_{p,mix}$) of the $H_2O/N_2/Cs$ gas mixture.

Similar to the specific heat capacity at constant pressure, to the specific gas constant, and to the adiabatic index; the speed of sound is independent of the pressure. Thus, there is a single response curve for each of these properties as a function of the cesium mole fraction (rather than one curve for each pressure value).

### C. ELECTRIC CONDUCTIVITY

While the previous subsections dealt with thermochemical properties of the preparatory $H_2O/N_2/Cs$ gas mixture that becomes plasma upon ionization; the current subsection deals with an electric property of the plasma, which is the electric conductivity.

Unlike the former thermochemical properties; the electric conductivity depends on the pressure, and thus a response curve is obtained for it at each pressure value.

Another distinction between the former thermochemical properties and the electric conductivity is that the latter exhibits a peak value at some level of cesium seed, rather than exhibiting a monotonic increase or decrease. The decline in the electric conductivity at a certain high level of ionization seed cesium might appear strange at first, given that cesium is the driver for this electric conductivity (it is the source of liberated free electrons in the plasma). However, this should not be surprising, because the presence of excessive amounts of electrons and cesium ions has a negative effect on the electric conductivity, and this effect becomes significant enough at a high cesium level (while remaining relatively small at low cesium levels) to cause the electric conductivity to decrease after it was increasing. This effect is the scattering of electrons by other electrons and by cesium ions (Coulomb scattering of charged particles), which suppresses the electrons' mobility and thus reduces the ability of the plasma to conduct a flow of electric current internally; and therefore while the electric conductivity initially increases with the seed level, this increase starts to slow down and reaches a peak value and then starts to decline as the seed level is increased further [216], [217], [218], [219], [220], [221], [222].

Fig. 5 shows three response curves of the hydrogen-based plasma as the cesium level increases, at the three selected absolute pressures (low: 0.0625 atm, moderate: 1 atm, high: 16 atm). Reducing the pressure has a favorable impact on the electric conductivity, and vice versa. Despite the similar overall profiles of the three curves, they are not simply related by linear scaling (multiplication by a constant). The peak electric conductivity for either of the three pressure values occurs (within a resolution of 1% change in cesium mole fraction) at the same cesium mole fraction of 6% (0.06). We found that this is the level of cesium seeding that optimizes the electric conductivity of this type of hydrogen-air plasma. The values of the maximum obtained electric conductivity for each pressure are enclosed within boxes for easy identification. These values are 56.0605 S/m at 0.0625 atm, 17.8321 S/m at 1 atm, and 4.7371 S/m at 16 atm. Thus, their relative ratio is 3.14380:1:0.26565, respectively.

Although increasing the pressure at a given cesium mole fraction means that the partial pressure of cesium increases (thus, more cesium atoms are available for thermal ionization), the intensified collisions with the increased number of molecules/atoms, and also the increased frequency of

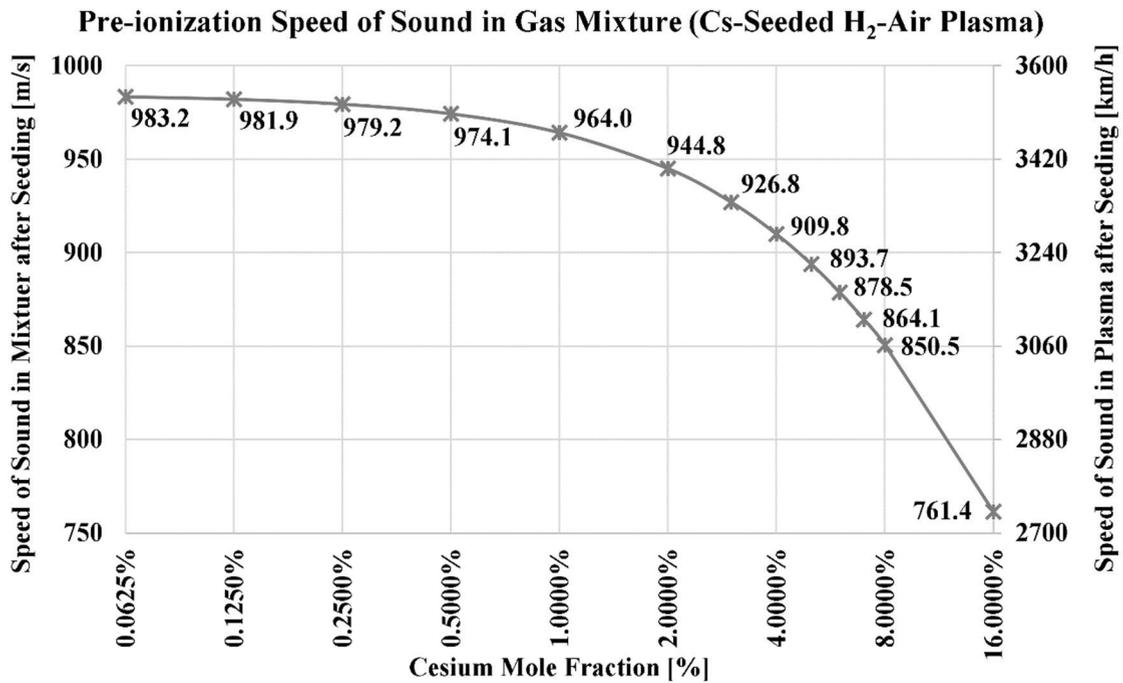

**FIGURE 4.** Response curve of the approximated speed of sound ($a$) in the hydrogen-based plasma.

Coulomb scattering due to having more charged particles both lead to an overall decline in the electric conductivity as the pressure of the cesium-seeded gas mixture increases which is accompanied with a decrease in the mean free path of free electrons (with free electrons being the effective faster charge carriers compared to the much heavier ions) within the plasms, thus poorer ability to conduct electric charge [223], [224], [225], [226], [227], [228], [229], [230].

When compared with a typical electric conductivity of seawater (taken as 5 S/m), the atmospheric pressure of hydrogen-based plasma can be below or above this level, depending on the cesium seed level; but in all the seed levels considered except the smallest one of $X_{Cs} = 0.0625\%$, the plasma electric conductivity exceeds that reference value, reaching 3.5664 times that value at the optimum $X_{CS} = 6\%$. At the low-pressure cases, the plasma electric conductivity well exceeds the seawater reference value under all considered seed levels; reaching 11.2121 times that reference value at the optimum $X_{CS} = 6\%$. In the high-pressure cases, the plasma electric conductivity is always below the reference seawater value, with the peak plasma electric conductivity approaching that reference seawater value, reaching a large fraction (0.9474) of it at the optimum $X_{CS} = 6\%$.

### D. VOLUMETRIC ELECTRIC POWER DENSITY

After all the elements needed to estimate the maximum volumetric electric power output to an electric load from a unit volume (1 m³) of hydrogen-based plasma in a magnetohydrodynamic (MHD) channel with Mach number 2 and magnetic flux density 5 T, this estimated theoretical upper limit (idealized operation) of the output power is shown in Fig. 6 as three response curves versus the cesium mole fraction at the three considered pressures.

While there is a similarity in these response curves with those previously presented for the electric conductivity in terms of having a mound-type profile with a single peak; this peak (regardless of the pressure value) occurs at a lower level of cesium ($X_{CS} = 3\%$ or 0.03, with a resolution of 1%) compared to the peak identified earlier in the case of electric conductivity ($X_{CS} = 6\%$ or 0.06, with a resolution of 1%). The peak values are highlighted within enclosing boxes. This shift in the peak (optimum seed level) is explained by the added influence of the speed of sound. Because the speed of sound is a strictly decreasing function of the cesium concentration, it is thus responsible for favoring a lower cesium concentration for maximizing the power output.

As expected, lower pressures correspond to higher powers (due to the higher electric conductivity). One cubic meter of the supersonic plasma is estimated to supply electric power of more than 1 GW (more precisely 1146.828 MW; comparable to typical capacities of commercial power plants) if the magnetohydrodynamic channel is allowed to operate at a low pressure of about 6% of the normal atmospheric pressure [231], [232], [233], [234], [235], [236]. This is an enormous density of power generation. Even if the actual power generation is a modest fraction of this theoretical estimate, it remains compelling, particularly since there are no moving mechanical parts in the MHD channel; thus, there is no need for mechanical maintenance and the matter of reliability due to mechanical failures is not applicable; and also the process is environmentally clean with no direct (scope 1) $CO_2$ emissions [237], [238], [239], [240], [241].

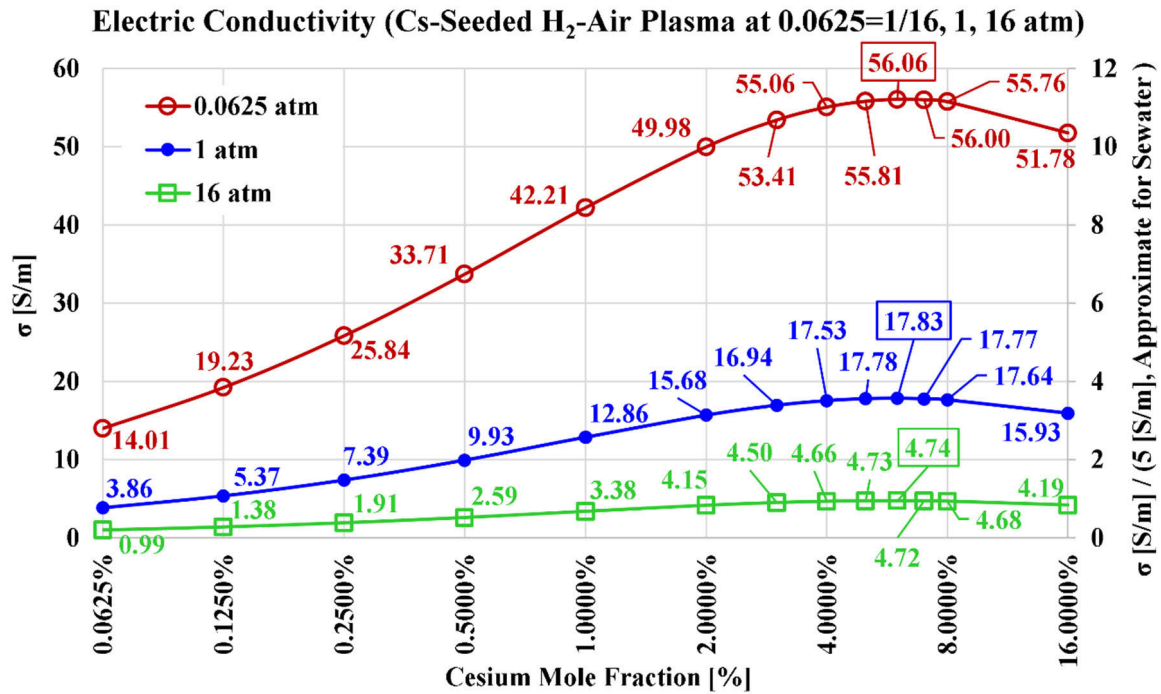

FIGURE 5. Response curves of the estimated electric conductivity ($\sigma$) in the hydrogen-based plasma.

If the pressure increases to a normal level of 1 atm, the peak power density drops by 68.28% to 363.771 MW/m$^3$; and if the pressure increases to a high level of 16 atm, the peak power density drops further by 73.45% to 96.591 MW/m$^3$. Thus, the relative ratio of the three peak power densities is 3.15261:1:0.26553, respectively.

Like the electric conductivity response curve, the three power densities are not scaled by constant factors, although they appear visually as such.

## V. DISCUSSION

In this section, we make four comments regarding the preceding results.

First, while we fixed the Mach number and the magnetic flux density at reasonable selected values in our study, there is no loss of generality incurred by freezing these two parameters. Equations (3, 4) justify this point where these two parameters are absorbed into the multiplicative constant in the expression used to compute the volumetric power density. They do not affect other important plasma properties such as the electric conductivity. Thus, changing the values of either of these two parameters merely causes vertical amplification or attenuation of the response curves of the volumetric power density.

Second, due to implied assumptions in our study (such as idealized combustion and uniform temperature), the reported power outputs should be regarded as estimated theoretical upper limits. The adoption of assumptions is inevitable to render the study manageable and focused. For example, if multi-step combustion with additional intermediate byproduct species are permitted, such as nitrogen oxides ($NO_x$); more parameters appear and they cause distraction in the study away from its main objectives. New questions arise and decisions need to be taken as the problem complexity is unnecessarily expanded and more specific details are covered; such as: What reaction mechanism to be used? What happens if other reaction kinetics are used? What fraction of argon or water vapor should be included in the modeled air? What happens if these fractions are varied? What coefficients should be used for the additional species when computing their electric effects? [242], [243], [244], [245], [246], [247], [248], [249], [250], [251], [252], [253], [254], [255], [256], [257], [258], [259], [260], [261] By eliminating these sources of uncertainty and variability, the results become more meaningful; with only two quantities allowed to vary, the cesium level and the pressure.

Third, to compare the potential volumetric power density of hydrogen-fueled magnetohydrodynamic (MHD) channels to automotive internal combustion engines (ICEs), the example of the gasoline engine in the Toyota RAV4 compact SUV vehicle is used here [262], [263]. That engine (engine code A25A-FKS) has a total displacement (the useful combustion volume) of 2487 cm$^3$ (2.5 liters), distributed over 4 engine cylinders, and can deliver a maximum power of 151 kW (at 6600 rpm, or revolutions per minute). Therefore, the peak volumetric power density is 151/2487=0.0607 kW/cm$^3$ or 60.7 W/cm$^3$. On the other hand, the peak volumetric power output obtained in our study with the normal pressure of 1 atm is 363.771 MW/m³ or 363.771 W/cm³, or almost six times that of the example internal combustion engine (ICE).

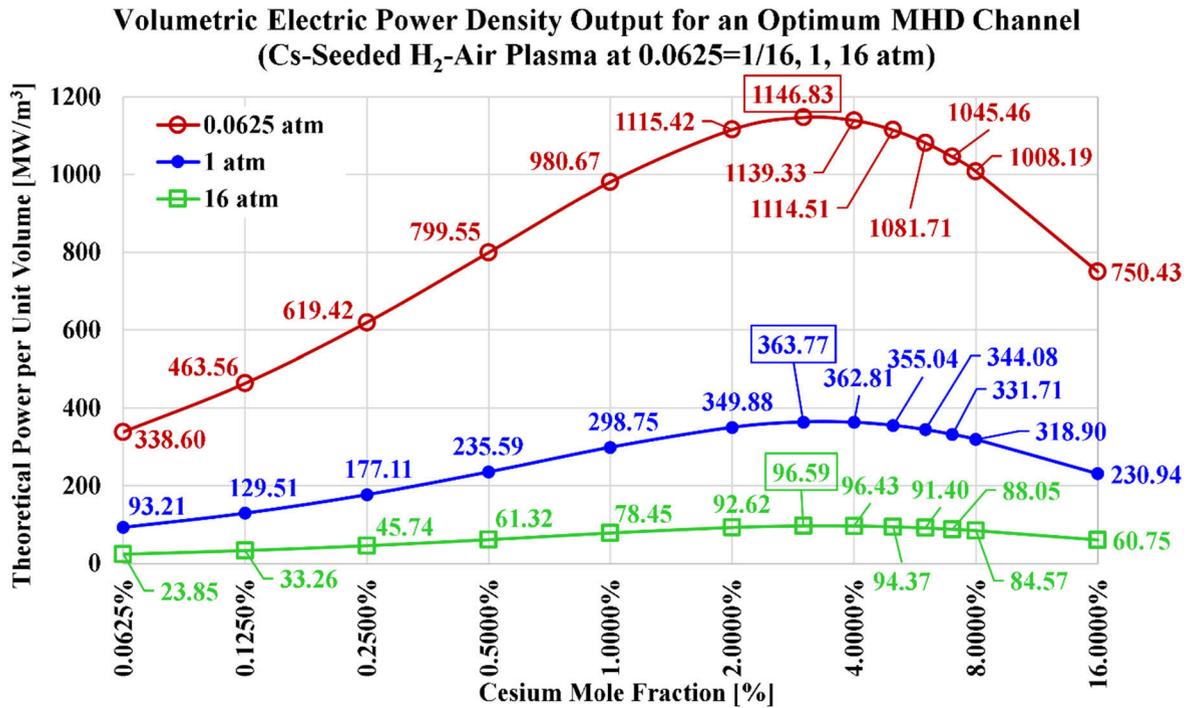

**FIGURE 6.** Response curves of the idealized power density ($P_V$) from hydrogen-based plasma.

Fourth, material compatibility is an important subject with regard to the successful commercialization of MHD-DPE. Extremely-high temperatures largely improve the electric conductivity of plasma, and thus increase the electric power output from an MHD generator. However, such a harsh environment may require special materials for the MHD plasma channel walls, as well as measures to ensure that electrodes remain functionally clean (electrically accessible to the plasma gas) despite the continuous exposure to flue gases, which may contain soot or slag-forming components.

## VI. CONCLUSION

In this section, we make three comments regarding the preceding results.

In the current study, we explored the utilization of hydrogen (particularly green hydrogen) in the concept of magnetohydrodynamic direct power extraction (MHD-DPE) through system-level modeling. As a multidisciplinary field, we applied principles of fluid mechanics, thermodynamics, physics, and chemistry to estimate thermochemical and electric properties of weakly-ionized equilibrium plasma formed by thermal ionization of cesium atoms seeded into water vapor and molecular nitrogen (as the products of idealized combustion of hydrogen in air). We studied the effect of the level of cesium (expressed as its mole fraction after seeding) on the performance of a hydrogen-based magnetohydrodynamic (MHD) channel. The cesium mole fraction was varied by two orders of magnitudes and various response curves were presented for three values of the pressure that also spanned two orders of magnitudes. The following remarks can be made:

- The addition of cesium to the water-vapor/nitrogen mixture has a limited effect on the adiabatic index, but the gas constant and speed of sound drop. The relative drop in the speed of sound relative to the original unseeded mixture is within 2% at a cesium mole fraction of 1%, and it is 4% at a cesium mole fraction of 2%.
- There is an optimum level of cesium seed for maximum electric conductivity of the plasma formed by water-vapor/nitrogen/cesium, which is approximately 6% (0.06). Further addition of seed harmfully reduces the plasma electric conductivity.
- There is an optimum level of cesium seed for attaining the maximum electric power output of the MHD channel, which is approximately 3% (0.03). Further addition of seed harmfully reduces the power output.
- Reducing the operating pressure of the MHD channel is influential in boosting the power output and the electric conductivity.
- Hydrogen-based MHD-DPE enables extremely high power generation densities, with the electric generation capacity of an entire conventional power plant may be achieved with a small space at the order of only one cubic meter.

## DECLARATION OF COMPETING INTERESTS STATEMENT

The author declares that they have no known competing financial interests or personal relationships that could have appeared to influence the work reported in this article.

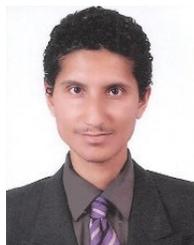

**OSAMA A. MARZOUK** received the B.S. and M.S. degrees in aerospace engineering and the Ph.D. degree in engineering mechanics.

Then, he was a Research Engineer in magnetohydrodynamic power generation and hydrogen membrane separation. Later, he moved to academia and is currently an Assistant Professor of mechanical engineering with Oman.

Dr. Marzouk is a member of the American Institute of Aeronautics and Astronautics (AIAA), the Association of Energy Engineers (AEE), the International Solar Energy Society (ISES), and the Life Cycle Initiative hosted by the United Nations Environment Program (LCI-UNEP). He was a recipient of the Who's Who in the World Recognition 2020 by Marquis Who's Who. He received the Certificate of Merit for a presented paper (co-authored with another colleague) from the International Association of Engineers (IAENG), in 2011.